\documentclass[%
 reprint,
 amsmath,amssymb,
 aps,
]{revtex4-2}

\usepackage[ansinew]{inputenc}
\usepackage[english]{babel}

\usepackage{mathrsfs}
\usepackage{xcolor}

\usepackage{graphicx}
\usepackage{dcolumn}
\usepackage{bm}

\DeclareMathOperator{\Tr}{Tr}

\begin{document}

\title{Stability of Rotating, Charged Fluids: Generalization of the H{\o}iland Conditions \\ in Newtonian Non-conductive Case}

\author{Kris Schroven}
\email{k.schroven@t-online.de}
 
\author{Vladim\'{\i}r Karas}

\author{Ji\v{r}\'{\i} Hor\'ak}%
\affiliation{Astronomical Institute, Czech Academy of Sciences, Bo\v{c}n\'{\i} II 1401, 14100 Prague, Czech Republic}%

\author{Audrey Trova}
\email{audrey.trova@zarm.uni-bremen.de}


\author{Eva Hackmann}
\affiliation{University of Bremen, Center of Applied Space Technology and Microgravity (ZARM),
28359 Bremen, Germany}%

\date{\today}

\begin{abstract}
We study the conditions for stability of electrically charged, non-conductive perfect fluid tori with respect to linear perturbations. To this end we employ Lagrangian perturbation formalism and we assume a system where the fluid orbits a central body. Gravitational field of the latter is described in the Newtonian framework. We first formulate the criteria valid for a general, non-axisymmetric situation, and then we concentrate on the axisymmetric model in more detail. In the latter case we generalize the H{\o}iland criterion of stability to non-vanishing electric charge and classify special examples. Toroidal structures with constant angular momentum distribution are found to be linearly stable. Subsequently, like in the uncharged case, rotating charged fluids are found to be unstable with respect to non-axisymmetric perturbations.
\end{abstract}


\maketitle


\section{\label{intro}Introduction\protect}
{ Equilibrium configurations of a rotating fluid and the conditions for their stability play an important role in astrophysics \cite{Tassoul,2003safd.book...75M}. Previously, we discussed a special case of} electrically charged, non-conductive, energetically bound figures of fluid equilibria around black holes \citep{chargedtori,Slan2013,schroven2018,trova2018,trova2020,kovar08, 2020CQGra..37x5007K}. A possibility of (even small) non-vanishing electrical charge greatly enlarges the parameter space of the system. The astrophysical motivation to explore the equilibrium structures of electrically charged, orbiting fluid arises from the fact that combined action of irradiation and rapid rotation are expected to lead to electric currents and charge separation. Unfortunately, no single approach is adequate to tackle in full generality the complex medium that emerges in these circumstances (however, see e.g.\ reference \citep{2010icp..book.....B} for discussion). We adopt the fluid description in terms of magnetohydrodynamical scenario, nonetheless, even in this framework the introduction of electromagnetic terms requires additional simplifying assumptions about their microphysical origin \citep{kovar14,kovar16}. A globally neutral astrophysical system can develop separated regions with non-zero net charge of the fluid which occur in vast range of systems from planetary neighborhood to accreting black holes. Numerical integration of the equations of motion is probably the only way to solve these structures under astrophysically realistic circumstances, but it often fails to give full understanding. On the other hand, analytical techniques are restricted to highly symmetrical configurations. Here we adopt the latter approach and we confine ourselves to axially and symmetrical, stationary perfect fluids.
		
A follow-up question arises: under which conditions the equilibrium solutions are stable. In this paper we will adopt the traditional approach to the stability analysis in the linear regime, which was applied previously to the neutral fluid and ideal magnetohydrodynamic equations (see e.g. \cite{Tassoul, FriedmanSchutza, friedman1975, MRIbalbus, Glampedakis2007}). This resulted in the  H\o iland equation. Here, we use the method of a linear Lagrangian perturbation theory to set up general stability equations. We can show, that like in the uncharged case, rotating charged fluids are unstable with respect to non-axisymmetric perturbations. For axisymmetric perturbations we can formulate a corresponding generalized version of the H\o iland criterion relevant to our specific setup of a charged, non-conductive fluid in an electromagnetic background field. 

The Lagrangian perturbation approach will be introduced in section
\ref{Lptheory}. In section \ref{cfequation} we will introduce the fluid equations for a non-conductive, charged fluid immersed in an electromagnetic background field. The Lagrangian approach is then used in section
\ref{sdce} to set up the general linear stability equations for the charged fluid and the instability of rotating, charged fluids with regards to non-axisymmetric perturbations is discussed briefly. The generalized  H\o iland conditions for a charged, non-conductive fluid are derived in section \ref{sdaxs}. These generalized conditions are afterwards discussed for selected special cases of model parameters, like angular momentum distributions and background fields, which were discussed in previous works concerning the existence of bound, charged fluid structures. A conclusion is given in section \ref{conclusion}.

\section{Lagrangian perturbation approach}\label{Lptheory}

While Eulerian perturbation theory describes the change of fluid variables at a fixed point in space, the Lagrangian approach relates fluid variables in equilibrium to the perturbed ones by a Lagrangian displacement vector field. This results in a description of the change of the fluid variables with respect to a frame dragged along with the displacement \cite{FriedmanSchutza}. The Lagrangian formalism was introduced by \citet{Lebovitz1961} and later on widely used to discuss the stability of (magnetohydrodynamic) fluids in a Newtonian and general relativistic setup (e.g. \cite{Lynden1967,FriedmanSchutza, Chandrasekhar1972,friedman1975,Glampedakis2007,Friedman2011}).

We denote the Eulerian physical coordinate by $x$, and the Lagrangian initial coordinate by $r$. In Lagrangian perturbation theory, the dynamical variable is the Lagrangian displacement field $\xi(r,t)$,
arising from deviations of a fluid line from its original course due to small perturbations of the fluid

\begin{align}
    x(t):=r+\xi(r,t).
\end{align}
It is worth mentioning that $\xi(r,t)=0$ leads to $r$ be the same as the usual comoving coordinate at initial time $t=0$.

The Lagrangian perturbation approach describes the perturbation of a fluid element with respect to a \glqq Lagrangian frame\grqq -- a frame that is embedded in the fluid and dragged along the perturbation direction. 
The Lagrangian perturbation $\Delta Q$ of an arbitrary tensor quantity $Q$ is then related to the Eulerian perturbation $\delta Q$ by \cite{FriedmanSchutza}:
\begin{equation}
    {\Delta Q=\delta Q+\mathcal{L}_\xi Q\,} ,
\end{equation}
where the Lie derivative $\mathcal{L}_\xi$ is given by 
	
\begin{equation}
	\mathcal{L}_\xi f=\xi^j \nabla_j f,
\end{equation}
for scalar fields $f$ and by
\begin{align}
	\mathcal{L}_\xi u^i= \xi^j \nabla_j u^i - u^j \nabla_j \xi^i,\, \\
	\mathcal{L}_\xi u_i= \xi^j \nabla_j u_i + u_j \nabla_i \xi^j.
\end{align}
for contra- and co-variant vector fields $u^i$, $u_i$. Here, the Einstein summation notation is used.
    
In order to express the conservation of mass and equation of state in terms of Lagrangian displacement following \cite{FriedmanSchutza} we consider 
\begin{align}
\label{LPv}
   &\Delta u^i =\partial_t \xi^i,\\
 &\Delta \rho = -\rho  \nabla_i \xi^i,\,  
\end{align}	
where $u^i$ is the fluid velocity and $\rho$ is the density. In addition, by utilizing the adiabatic assumption, as well as a polytropic equation of state, one finds that
\begin{align}
 \Delta p &= \frac{\gamma p}{\rho}\Delta \rho & &\text{ and } & \Delta S = 0\; .   
 \label{LPpS}
\end{align}
Here $p$ is the pressure and \textbf{$\gamma=(\partial\ln p/ \partial\ln\rho )_S$} 
is the adiabatic index,  while $S$ is the entropy of the fluid.
	
\subsection{Canonical energy and canonical displacements}
An important fundamental property of the Lagrangian perturbation approach is gauge freedom
related to the displacement vector field $\xi$. In fact, there is a class of \textit{trivial displacements} $\eta$, for which the corresponding Eulerian change in pressure, density, velocity and entropy of fluid vanishes.
As a result, two canonical displacements $\xi^i$ and $\hat\xi^i$ correspond to the same physical perturbation if they differ only by a trivial displacement:
\begin{align}
    \xi^i =\hat{\xi}^i + \eta^i\;.
\end{align}
However, there is a subtlety here. Even though $\xi$ and $\hat{\xi}$ describe identical physical perturbations, the canonical energy $E_c$ is not preserved
\begin{align}\label{problemofLD}
    E_c(\xi^i)\neq E_c(\hat{\xi}^i + \eta^i).
\end{align}
The canonical energy is defined in terms of the symplectic structure $W(\xi, \partial_t \xi)$ of a Lagrangian displacement and its time derivative \cite{FriedmanSchutza}

\begin{align}
    \label{canonie}
    E_c:= \frac{1}{2} W(\xi, \partial_t \xi) = \frac{1}{2}H(\xi,\xi)\, ,
\end{align}

The canonical energy plays a key role in the formulation of the stability criterion for a fluid configuration in the following sense \cite{FriedmanSchutzb}: \newline\newline
The configuration is
\begin{enumerate}
    \item dynamically stable, if $\quad \, E_c(\xi)>0$,
    \item dynamically unstable, if $E_c(\xi)=0$,
    \item secularly unstable, if $\quad \;\; E_c(\xi)<0$
    
\end{enumerate}
for all initial conditions (in the last case, in particular, for non-axisymmetric conditions).

{This classification has a straightforward interpretation: 
if the canonical energy is positive for all possible configurations, the perturbation can only die away over time, while $E_c$ is decreasing. In the case that the canonical energy becomes negative for any possible configuration, the perturbations does not die away, as $E_c$ can only decrease and the system is dynamically unstable.}

We restrict the Lagrangian displacements to a subclass of so-called \textit{canonical displacements}, which are orthogonal to all trivial displacements with respect to the symplectic product,

\begin{align}
    W(\xi,\eta)=0, \quad \forall \eta, 
\end{align}
where $\eta$ is the trivial displacement \cite{FriedmanSchutza}. Therefore, for this subclass of canonical displacements the canonical energy is invariant.

\section{A charged, non-conductive fluid in the presence of an electromagnetic field}
\label{cfequation}
A charged, non-conductive fluid in the vicinity of an electromagnetic background field is described by the following set of equations; the Euler equation of a charged, non-conductive fluid and the continuity equations for mass and charge, respectively
	
\begin{align}
	\frac{D u_a}{Dt}+\frac{\nabla_a p}{\rho} + \nabla_a \Phi_G -\frac{q}{\rho}\left(F_{ab}u^b-\nabla_a\Phi_E\right)=0 \label{feq}\; ,\\
	\partial_t{\rho}+\nabla_b(\rho u^b)=0 \label{contm}\; ,\\
		\partial_t{q}+\nabla_b(q u^b)=0 \label{contch}\; .
\end{align}
Here, the external gravitational and electrostatic field are described by their potentials $\Phi_\mathrm{G}$ and $\Phi_\mathrm{E}$, respectively. Magnetic field is represented by the antisymmetric Maxwell tensor $F_{ab}$ ($F_{ab} = \nabla_a B_b - \nabla_b B_a$) and $\rho$ and $q$ are mass and charge densities.
Each term in Eq. \eqref{feq} represents different force, that arises due to internal pressure gradients, gravity of the central compact body, and the electromagnetic forces acting on the charged fluid medium, respectively \cite{rezzolla13, Slan2013}.
Finally, we introduced the total time derivative
\begin{align}
    \frac{D}{Dt}:=\partial_t+u^j\nabla_j
\end{align} 
in Eq. \eqref{feq}.
In the present discussion we neglect the fluid self-gravity and the electric current self-interaction. 
For simplicity reasons we introduce the specific charge
\begin{align}
     \frac{q}{\rho}:=\hat q\;.
 \end{align}
From equations (\ref{contm}) and (\ref{contch}) we find that
\begin{align}
	\label{qcons}
    \frac{D}{Dt} \hat q=0\, ,
\end{align}
which basically describes that the specific charge is preserved along the fluid course (i.e. the specific charge is a material invariant).
 
 Within the setup of our charged, non-conductive fluid we will use the Lagrangian perturbation approach, described above. Next to the expressions for the Lagrangian perturbation of the fluid quantities pressure, mass density, entropy and fluid velocity, already introduced in \cite{FriedmanSchutza} (see Eqs. \eqref{LPv}--\eqref{LPpS}), we receive for the Lagrangian perturbation of the specific charge 
\begin{align}
	\Delta \hat q &=0 \, .
\end{align}
Additionally, we derive the orthogonality condition between the canonical and trivial displacements in our charged setup, by closely following the procedure used in \cite{FriedmanSchutza}:
\begin{equation}
    \varepsilon^{ijk}\nabla_i\hat q\nabla_j
    \left(\Delta_{\hat \xi} u_k-\hat q F_{kb}{\hat \xi}^b \right)=0\; ,
	\label{canonD}
\end{equation}
where $\Delta_{\hat{\xi}}$ is the Lagrangian perturbation related to the displacement $\hat{\xi}$.
Our stability discussion will be restricted to the subclass of canonical displacements $\hat \xi^i$ that satisfy Eq. \eqref{canonD}.
	
\section{Stability discussion with the help of the canonical Energy}	\label{sdce}
It is our aim in this section to first derive the canonical energy of the perturbed charged fluid system, and afterwards use it for further stability discussions, by following the approach used in \cite{FriedmanSchutza}. We are only interested in perturbing the fluid and not the external gravitational or electromagnetic field.
Therefore, we assume that the Eulerian perturbations of these external fields vanish, $\delta F_{ab}=\delta \nabla_a \Phi_G=\delta \nabla_a \Phi_E=0$. Furthermore, we assume, that all discussed systems have a vanishing charge and mass density at infinity. With these assumptions, the Lagrangian perturbation of the Euler equation \eqref{feq} is given by
\begin{align}	\label{lp1}
	\begin{split}
	\Delta_\mathrm{E}= &\partial^2_t \xi_a + \left(2g_{ab}u^c\nabla_c-\hat q F_{ab} \right) \partial_t \xi^b+\left(u^b\nabla_b\right)^2 \xi_a\\
	&+\frac{1}{\rho}\left(\nabla_b\xi^b\nabla_a p-\nabla_a \xi^b \nabla_b p-\nabla_a(\gamma p \nabla_b\xi^b)\right)\\
	&+\xi^b\nabla_b\nabla_a\Phi_G+\hat q\xi^b\nabla_b\nabla_a\Phi_E\\
	&-\hat q \left(F_{ab}u^\gamma\nabla_\gamma\xi^b+\xi^b\nabla_bF_{a\gamma}u^\gamma\right)=0.\
\end{split}
\end{align}	
Without electric and magnetic fields, this result agrees with the previous works. However even with them, the equation is of the form
\begin{align}
    A_{ab} \partial^2_t \xi^b + B_{ab} \partial_t \xi^b +C_{ab} \xi^b=0\, .
\end{align}

If $A_{ab}$, $B_{ab}$ and $C_{ab}$ are hermitian, anti-hermitian and hermitian, respectively (see Appendix \ref{hermF} for proof), Eq.\ \eqref{lp1} can be derived from a variational principle with the action potential 
	\begin{equation}
	\begin{split}
   I: =&\int \mathscr{L} \,dV\\
	   =& \frac{1}{2}\int \dot \xi^{a} A_{ab}\, \dot \xi^b + \dot \xi^{a} \,B_{ab}\, \xi^b -\xi^{a} \,C_{ab}\, \xi^b  \,dV\,.
	\end{split}
 \label{Iintegral}
	\end{equation}

We can now calculate the canonical energy $E_c$ of the perturbed system, using the Lagrangian $\mathscr{L}$.  From \eqref{canonie} we obtain

	\begin{align}  
 E_c=\frac{1}{2}H(\xi_a,\xi^a)&=\int \dot \xi^{a} \left( \frac{\partial \mathscr{L}}{\partial \dot\xi^{a}} -  \mathscr{L}\right)\;dV 
	\end{align}
By using Eq. \eqref{lp1}, the expression for the canonical energy in our set-up is explicitly given by
\begin{widetext}
	\begin{align}
	    \label{canE}
	    \begin{split}
      E_c=\frac{1}{2}\int&\left\{\rho\, \lvert \dot\xi\rvert^2 -  \rho\lvert u^\gamma \nabla_\gamma\xi\rvert^2+\gamma p \lvert\nabla_a\xi^a\rvert^2+{\xi^a}^\ast \nabla_a p\nabla_b\xi^b+\xi^b \nabla_b p \nabla_a {\xi^a}^\ast+{\xi^a}^\ast\xi^b\rho\left[\nabla_b\nabla_a p+\nabla_b\nabla_a\Phi_G\right]\right.\\
	    &\left.+{\xi^a}^\ast \xi^b\rho\left[\hat q\nabla_b\nabla_a\Phi_E-\frac{\hat q}{2}\left(\nabla_aF_{b\gamma}+\nabla_bF_{a\gamma} \right)u^\gamma\right]
	    -\rho\frac{\hat q}{2} \left({\xi^a}^\ast F_{ab}u^\gamma\nabla_\gamma\xi^b +\xi^a F_{ab}u^\gamma\nabla_\gamma{\xi^b}^\ast \right) \right\}dV\; ,
	    \end{split}
	\end{align}
\end{widetext}
where we used $ \nabla_a(\rho u^a)=0$ and assumed that the mass density, charge density and pressure of the fluid vanishes at the surface of the considered system.


Equation \eqref{canE} could be used to check the stability of the normal modes of the system, where the normal mode under discussion is represented by a specific $\xi^a$.

\subsection{Instability with regards to non-axisymmetric perturbations}

The classical rotating, isentropic or nonisentropic, uncharged fluid is unstable with regards to non-axisymmetric perturbations \cite{FriedmanSchutzb}. Perturbation modes of negative energy transport angular momentum and energy to infinity, while the rotating fluid is slowing down. 
The energy transfer is inspired by variation of various multipoles 
caused by the perturbations. This instability, however, is irrelevant for slowly rotating fluids \cite{FriedmanSchutzb}. 
	    
The discussion by  \citet{FriedmanSchutzb} can be extended to our case of a charged rotating fluid, leading to the same instability. To check this, we use the following ansatz for the non-axisymmetric perturbation of the fluid in cylindrical coordinates:
	    \begin{align}
	        \xi^a=&\zeta^a(r,z) \mathrm{e}^{im\phi}, &  \dot \xi^a=&\dot \zeta^a(r,z) \mathrm {e}^{im\phi}\; .
	    \end{align}
The ``amplitudes'' $\zeta^a(r,z)$ and $\dot \zeta^a (r,z)$ can also depend on $m$ but should be bounded everywhere, independent of $m$. They are chosen such that $\zeta^\phi$ vanishes and condition \eqref{canonD} is fulfilled
	    \begin{align}
	        \zeta_a&= f\nabla_a\hat q\, , & \dot \zeta_a &=\frac{2}{im} \partial_a\left(r\Omega\zeta_r-F_{b\phi}\zeta^b\right)\ ,
	    \end{align}
	   where $f=f(r,z)$ has to be a smooth function, which is constant along the fluid trajectories $(\partial_t+\mathcal{L}_u)f=0$. Accordingly, the divergence of $\xi^a$ becomes
	    \begin{align}
	        \nabla_a\xi^a=\nabla_a\zeta^a(r,z) \mathrm{e}^{im\phi}\; .
	    \end{align}
	    
	    Plugging the ansatz into Eq.\ \eqref{canE} leads to
	    \begin{align}
	    \label{pertE}
	    \begin{split}
	        E_c=K&-\int\rho\hat q\Omega\; \zeta^a\zeta^b \left( \nabla_a\nabla_bA_\phi-\nabla_a\nabla_b \Phi_E \right)\,dV\\
	        &-m^2\int{\rho \Omega^2 \lvert\zeta\rvert^2 \,dV}\, .
	   \end{split}
	    \end{align}
	    Here, $K$ is a term that includes all terms from perturbing the uncharged fluid with the given ansatz, which are independent from $m$. { According to the named assumptions, both $K$ and the first integral in Eq.\ \eqref{pertE}, which contains the electromagnetic effects and is independent of $m$ as well, are bounded from above.} Therefore $E_c$ will eventually become negative for sufficiently high frequencies in the non-axisymmetric perturbation.
     
     {
     As one can see from Eq.\ \eqref{pertE}, the electromagnetic interactions of the fluid do not contribute any term that would depend on the  perturbation frequency $m$. Following the discussion in \citet{FriedmanSchutzb}, we assume that a set of non-axisymmetric perturbations exist that do not die away in time, therefore leading to instability.
     
     }
	\section{Axisymmetric perturbations: The generalization of the Solberg-H\o iland conditions}
	\label{sdaxs}
	    We saw that rotating, charged, non-conductive fluid structures will slow down over time under non-axisymmetric perturbations. We will now turn our focus on the special case of axisymmetric perturbations, that preserve the angular momentum of the system. In the non-charged fluid case, this stability discussion leads to the Solberg-H\o iland conditions or otherwise the Rayleigh criterion in the special case of fluids with an overall constant entropy $S$.
	    
	    Under certain conditions for the perturbations in the fluid, the first order variation $\delta^1 E_{\mathrm{tot}}$ of the total energy of a charged fluid vanishes (the solution to the Euler equation \eqref{feq} has an extremal total energy only with respect to this class of perturbations), and the canonical energy coincides with the second order variation $\delta^2 E_{\mathrm{tot}}$ of the total energy of the fluid. Since the variation $\bm{\delta} E_{\mathrm{tot}}$ of the total energy is a physical property of our configuration, $E_c$ will loose its gauge dependence under this conditions, and a discussion of canonical and trivial displacements is not necessary anymore. Perturbations of this class conserve the fluid mass and charge, leave the system adiabatic and further conserve the redefined fluid's angular momentum ($\Delta \hat u_\phi=\Delta(u_\phi+\hat q A_\phi)=0$; see appendix \ref{svar}). A conservation of mass and charge is automatically satisfied under any perturbations, if the continuity equation for mass and charge (see Eqs.\ \eqref{contm}--\eqref{contch}) are valid.
	    
	    In this chapter we will focus the discussion on adiabatic, axisymmetric perturbations that conserve the redefined angular momentum $\hat u_\phi$ of the fluid. In this case $E_c$ coincides with the second variation of the total energy of the system and we can follow the approach used by \citet{Tassoul}.
	
\subsection{Case of axial symmetry}
    Axissymmetry (independence of the angular coordinate $\phi$ corresponding to that symmetry) leads to a conservation constant. For an uncharged fluid, the conserved quantity is the angular momentum of the fluid $L=u_\phi$. In our case of a charged, non-conductive fluid, affected by an electromagnetic background field, the conserved quantity is a \textit{redefined angular momentum} $\hat L=\hat u_\phi$. It can be directly derived (in e.g. cylindrical or spherical coordinates) from the $\phi$-component of the charged Euler equation \eqref{feq}, while using Eq.  \eqref{qcons}:
	\begin{align}
	&\frac{D}{Dt} u_\phi-\hat q F_{\phi b}u^b=0 \, ,\\
	\Rightarrow\; 	&\frac{D}{Dt}\left(u_\phi+ \hat q A_\phi \right)=	\frac{D}{Dt}\hat u_\phi=0 \, ,
	\end{align}
	where we assumed, that the electromagnetic background field is stationary.
	
	Furthermore, the magnetic vector potential can be restricted to its $\phi$-component in our axisymmetric case ( $A_a=A^\varphi \eta_a$, while $\eta^a=\hat e_\varphi^a$ corresponds to the Killing vector for axial symmetry).
	
	In order to use the conservation of $\hat u_\phi$ in the following stability analysis, we will rewrite the charged Euler equation \eqref{feq} in terms of
	\begin{equation}
	\hat u_\mu=u_\mu +\hat q A_\mu\, .
	\label{Lred}
	\end{equation}
	It becomes:
	\begin{align}
	\begin{split}
	\left(\hat u^b-\hat q A^b \right)\nabla_b \left(\hat u_\mu-\hat q A_\mu \right)- \hat q F_{\mu b}\left(\hat u^b-\hat q A^b \right)&\\  +\underbrace{\frac{\nabla_\mu p}{\rho} + \nabla_\mu \Phi_G +\hat q \nabla_\mu\Phi_E}_{T_2}&=0\, ,
	\end{split}\\
	\begin{split}
	\hat u^b\nabla_b \hat u_\mu-\hat q\left[A^b\nabla_b \hat u_\mu +\hat u^b \nabla_b A_\mu \right] +\hat q^2 A^b\nabla_b A_\mu& \\
	-\hat q \left[\nabla_\mu A_b -\nabla_b A_\mu\right](\hat u^b-\hat q A^b)+T_2&=0\, , 
	\end{split}
	\\
	\hat u^b\nabla_b \hat u_\mu-\hat q\left[A^b\nabla_b \hat u_\mu +\hat u^b \nabla_\mu A_b \right] +\hat q^2 A^b\nabla_\mu A_b +T_2&=0\, .
	\end{align} 
	
	In the case of circular fluid motion and a magnetic vector field restricted to its $\phi$-component, the second term in the equation above reduced to:
	\begin{align}
	\label{sasFEQ}
	\begin{split}
	&\left[A^b\nabla_b \hat u_\mu +\hat u^b \nabla_\mu A_b\right]\\
	&=g_{\phi\phi} \hat u^\phi  {A^\phi}_{,\mu}+ {g_{\phi\phi}}_{,\mu} \hat u^\phi A^\phi
	-2\cdot\left(\frac{1}{2}{g_{\phi\phi}}_{,\mu} \hat u^\phi A^\phi\right),
	\end{split}\\
	&=g_{\phi\phi} \hat u^\phi  {A^\phi}_{,\mu}.
	\end{align} 
	Finally, we end up with the following new version of the Euler equation \eqref{feq} in terms of $\hat u^b$:
	\begin{align}
	\label{neweuler}
	\begin{split}
	\hat u^b\nabla_b \hat u_\mu-\hat q\left( \hat u_\phi  {A^\phi}_{,\mu}-\nabla_\mu\Phi_E\right)& \\
	+\frac{1}{2}\hat q^2 (\left|A_\phi \right|^2)_{,\mu}+\frac{\nabla_\mu p}{\rho} + \nabla_\mu \Phi_G &=0\, .
	\end{split}
	\end{align} 
	\subsection{Rayleigh criterion for the charged, non-conductive fluid}
	 We are only interested in the perturbation of the fluid, which is why we do not consider any perturbation of the external gravitational or electromagnetic field. Therefore, $\delta F_{ab}=\delta \nabla_a \Phi_G=\delta \nabla_a \Phi_E=0$. A linear perturbation of our new expression of the Euler equation \eqref{neweuler} leads in the case of an initially unperturbed fluid in circular motion to:
	 \begin{widetext}
	 \begin{align}
	 E_c=&\delta^2 E=\Delta_{\mathrm{E}} =\int \xi^a M_{ab} \xi^b\, dV \hspace{10.75cm} \nonumber \\
	 \label{mab}
	 \begin{split}
	 =&\int \xi^a\left[\frac{\nabla_b \hat L^2}{2g_{\varphi\varphi}^2}\partial_a g_{\varphi\varphi}+ \frac{\nabla_a p}{\rho}\left(\frac{\nabla_b \rho}{\rho} -\frac{\nabla_b p}{\gamma p}\right)
	 -\nabla_b \hat q \nabla_a\Phi_E -\frac{1}{2}\nabla_b \hat q^2 (\left|\textbf{A} \right|^2)_{,a}+\nabla_b (\hat q\hat L) {A^\phi}_{,a}\right]\xi^b  \rho dV
	 \\
	 &+\underset{\approx 0}{\underbrace{\int \frac{(\delta p)^2}{\gamma p} dV}}+ \underset{=0}{\underbrace{\int_{\partial V} \xi^a\delta p\, d\hat n_a}}\,.
	 \end{split}
	 \end{align}
	 \end{widetext}
	 where we assumed axisymmetric perturbations, which preserve the redefined angular momentum: $\Delta \hat L=0$. Furthermore, we neglect the integral over $(\delta p)^2$. This approximation is valid, if the characteristic time of the disturbance exceeds the travel time of a sound wave in the perturbed area. This approximation basically neglects any possible p--modes, that could arise due to the perturbation \cite{Tassoul}.
	 
	 In Eq. \eqref{mab}, the canonical energy is now expressed in form of a perturbation matrix $M_{ab}$. As long as $M_{ab}$ is positive definite, $E_c$ will be positive for all $\xi^a$, and the system can be considered stable. The two new stability conditions are therefore given by the conditions for a matrix being positive definite. According to \citet{Tassoul}, these conditions are: 
	\begin{align}
	\label{cposdef}
	&\Tr(M_{ab})= M^a_a\geq0, & \det \left(M_{ab}\right)&\geq0\, .
	\end{align}
	
	The pressure and density of a perfect fluid are connected to the entropy $S$ as $\alpha \nabla_aS=-\left(\frac{\nabla_a \rho}{\rho}-\frac{\nabla_a p}{\gamma p}\right)$ (see e.g. \cite{Tassoul}). By using the found expression for $M_{ab}$ in \eqref{mab}, the two stability conditions take the form:
	\begin{align}
		\label{trcon1}
		\begin{split}
		&\Tr\left(M_{ab}\right)\\
		&\quad= g^{ab}\left[\left(\hat q {A^\phi}_{,a}-{\left(\frac{1}{g_{\varphi\varphi}}\right)}_{,a}\hat L\right)\nabla_b \hat L-\alpha \frac{\nabla_a p}{\rho}\nabla_b S \right.\\
		&\left.\quad+\left(E_a - \hat q (\left|\textbf{A} \right|^2)_{,a}+\hat L {A^\phi}_{,a}\right)\nabla_b \hat q \right]\geq 0 \, ,
		\end{split}\\
		\label{detcon1}
		\begin{split}
		&\det \left(M_{ab}\right)\\
		&\quad= \varepsilon^{ab}k^1_ak^2_b\; \varepsilon^{\alpha\beta}\ell^1_\alpha\ell_\beta^2 +  \varepsilon^{ab}k^1_ak^3_b\; \varepsilon^{\alpha\beta}\ell^1_\alpha\ell_\beta^3 \\
		&\quad+  \varepsilon^{ab}k^2_ak^3_b\; \varepsilon^{\alpha\beta}\ell^2_\alpha\ell_\beta^3\geq 0\, ,
		\end{split}
	\end{align}
	with 
	\begin{align}
	k^1_a=&\hat q {A^\phi}_{,a}-{\left(\frac{1}{g_{\varphi\varphi}}\right)}_{,a}\hat L\, , & 	\ell^1_b=&\nabla_b \hat L\, , \\
	k^2_a=& \nabla_a p\, ,& \ell^2_b=& -\alpha\nabla_b S\, ,\\
	k^3_a=& E_a - \hat q (\left|\textbf{A} \right|^2)_{,a}+\hat L {A^\phi}_{,a}\, ,& \ell^3_b=& \nabla_b \hat q\, .
	\end{align}
	
	In the case of a vanishing electromagnetic field, the stability conditions \eqref{trcon1} and \eqref{detcon1} reduce to the Solberg-H\o iland conditions. If aditionally the entropy is constant, the stability conditions further reduce to the Rayleigh criterion:
	\begin{equation}
		\bm{\nabla}(\hat L^2)\geq 0\, .
	\end{equation}
	
\subsection{Specific cases}
\begin{figure}
    \centering
    \includegraphics{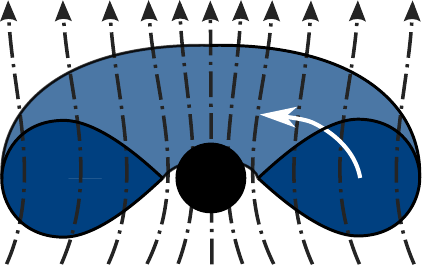}
    \caption{Exemplary depiction of a gravitationally bound, rotating fluid structure (blue) immersed in a background magnetic field (grey arrows aligned with the vertical direction). The central body (black circle) can be electrically charged, which adds an electric field to the system (not depicted here). The fluid circulates in the azimuthal direction around the central body with a given velocity distribution.}
    \label{fig1}
\end{figure}

    In this section we will apply the new found stability conditions for a non-conductive, charged fluid in an electromagnetic background field to several special cases with a given electromagnetic field and angular momentum distribution. We assume a barotropic equation of state for all cases, which leads to a constant entropy distribution throughout the fluid structure. All discussed special cases that follow were examined in previous works concerning the existence of bound charged fluid structures, either in a classical or general relativistic setup. This discussion below is meant to evaluate, if the found bound structures are stable concerning axisymmetric perturbations \cite{chargedtori,Slan2013,kovar14,kovar16,schroven2018,trova2018,trova2020}. An exemplary depiction of such bound fluid structures is presented in Fig. \ref{fig1}.
    
    \paragraph{Case I: Constant angular momentum, homogeneous magnetic field.}
    In this case, the cylindrical coordinate system is used. In this coordinate system, the vector potential for a homogeneous magnetic field is given by:
    \begin{align}
        A_\phi&=\frac{1}{2}B_z r^2,& &\text{and } & A^\phi&=\frac{1}{2}B_z\, .
    \end{align}
    We consider a magnetic field only, therefore setting $\Phi_E=0$. The two conditions for stable structures in case of a constant angular momentum distribution and under axisymmetric perturbations reduce in this set up to:
    \begin{align}
        \Tr(M_{ab})&=(\hat q B_z)^2 + \frac{L}{r^3}B_z \partial_r(\hat q r^2)\; ,\\
        \det(M_{ab})&=0\; .
    \end{align}
    According to this result, a charged fluid structure can be considered stable to axisymmetric perturbations, if the specific charge distribution -- or more precisely $\hat qr^2$ -- increases outwards in case of a repulsive Lorentz force or decreases outwards in case of an attractive Lorentz force.
    
    This is the direct generalization of the classical uncharged Polish doughnut set up with a constant angular momentum distribution. Since uncharged ideal fluid structures are considered stable, if the specific angular momentum in increasing outwards, a small fluid charge can stabilize or destabilize the structure. 
     \paragraph{Case II: Constant angular momentum, charged central object.}
    In this case, the spherical coordinate system is used. In this coordinate system, the electric potential arising from the charge of the central object  is given by:
    \begin{equation}
       E_a=\left(\frac{Q}{r^2},0,0\right)\; .
    \end{equation}
    We consider no magnetic field, therefore setting $A_a=0$. The two conditions for stable structures for a constant angular momentum distribution and under axisymmetric perturbations reduce in this set up to:
    \begin{align}
        \Tr(M_{ab})&=\partial_r {\hat q} E_r \; ,\\
        \det(M_{ab})&=0\; .
    \end{align}
    According to this result, a charged fluid structure can be considered stable to axisymmetric perturbations, if the charge increases (same charge as the BH) or decreases (opposite charge of the BH) outwards. This means, that in case of a repulsive electric force on the fluid, the repulsive force should fall off slower than the attractive gravitational force and vice versa for the case of an attractive electric force on the fluid.
    \paragraph{Case III: Constant angular momentum, charged central object immersed in a homogeneous magnetic field.}
    Considered is a homogeneous magnetic field and a central object with charge. The two conditions for stable structures for a constant angular momentum distribution and under axisymmetric perturbations reduce to:
    \begin{align}
        \Tr(M_{ab})=&\partial_a\hat q E^a+(\hat q B_z)^2 +LA^\phi g^{ab}\frac{\partial_ag_{\phi\phi}}{g_{\phi\phi}^2} \partial_b(g_{\phi\phi}\hat q ) \; ,\\
        \begin{split}
        \det(M_{ab})=&\hat q A^\phi\frac{\hat L}{g_{\phi\phi}^2}\; \left(g^{ab}\partial_a g_{\phi\phi}\partial_b g_{\phi\phi} E^c \partial_c \hat q \right.\\
        &\left.- \partial_a g_{\phi\phi}E^a g^{cd} \partial_c g_{\phi\phi} \partial_d \hat q\right) \; .
        \end{split}
    \end{align}
    In cylindrical coordinates, this leads to:
    \begin{align}
        \Tr(M_{ab})&=\partial_a\hat q E^a+(\hat q B_z)^2 + \frac{L}{r^3}B_z \partial_r(\hat q r^2) \; ,\\
        \det(M_{ab})&=\hat q B_z\; E^z \partial_z \hat q\, \frac{2 L}{r^2}+(\hat q B_z)^2\; E^z \partial_z \hat q \; .
    \end{align}
    
    Generally one can deduce form this, that the fluid structure is more likely to be stable, if only repulsive electromagnetic forces occur ($\hat q E^a>0$ and $\hat q B_z L>0$). At the same time the charge density of the fluid structure should increase outwards.
    

   \begin{figure*}[tbh!]
    \centering
    \includegraphics[width=\textwidth]{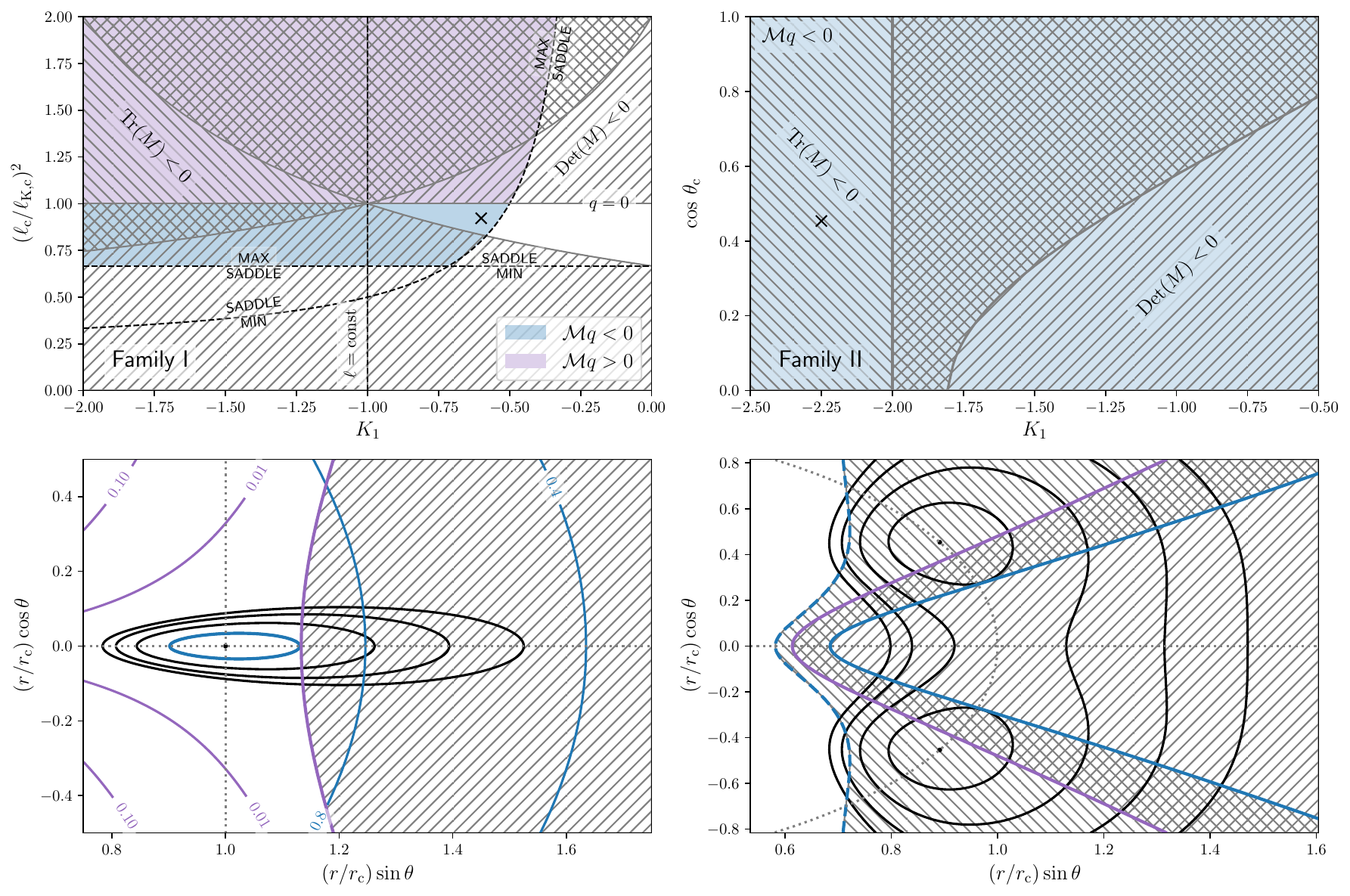}
    \caption{The figures of equilibrium and the corresponding distribution of pressure surfaces in charged slender tori immersed in the dipole magnetic field corresponding to two Families lying {\em in the equatorial plane} (I and II); their tentative existence was originally discussed by Slan\'{y} et.\ al \cite{Slan2013}. \textit{Top left panel:} These tori correspond to `Family I' solutions with the specific angular momentum distribution $\ell = \ell_\mathrm{c}[(r/r_\mathrm{c})\sin\theta]^{K_1}$. The horizontal axis shows the parameter $K_1$; the parameter $(\ell_\mathrm{c}/\ell_\mathrm{K,c})^2$ shown on the vertical axis measures how much the rotation of the torus deviates from the Keplerian value at the torus center due to Lorentz force; the value $(\ell_\mathrm{c}/\ell_\mathrm{K,c})^2$ corresponds to uncharged tori, values greater/smaller than one correspond respectively to positively/negatively charged tori. The dashed areas denote regions in the parameter space, where the criteria for stability are not satisfied. The graph demonstrates that stable configurations do exist, however they are considerably restricted by the stability criterion and limited to a small region of the parameter space. \textit{Bottom left panel:} Example of the equi-pressure surfaces of negatively charged tori of finite size, whose parameters correspond to the cross shown in the top panel. Even though the center of the torus is stable, our criteria limit significantly possible sizes of the tori. Out of the three configurations shown by the thick black solid line, only the innermost one corresponds to an entirely stable torus. The corresponding panels in the right column have been constructed in the way analogous to the left column but for the case of Family~II (off-equatorial tori). In the latter case, we do not find any closed equi-pressure surface that would enclose a region of stability.}
    \label{fig2}
\end{figure*}

    \paragraph{Case IV: Constant angular velocity, homogeneous magnetic field.}
    In this case, the cylindrical coordinate system is used. In this coordinate system, the vector potential for a homogeneous magnetic field is given by:
    \begin{align}
        A_\phi&=\frac{1}{2}B_z r^2,& &\text{and } & A^\phi&=\frac{1}{2}B_z\, .
    \end{align}
    We consider a magnetic field only, therefore setting $\Phi_E=0$. The two conditions for stable structures in case of a constant angular velocity distribution and under axisymmetric perturbations reduce in this set up to:
    \begin{align}
        \Tr(M_{ab})&=\left(2\Omega +\hat q B_z\right)^2+ \partial_r {\hat q} B_z \, r\Omega \; ,\\
        \det(M_{ab})&=0\; .
    \end{align}
    According to this result, a charged fluid structure can be considered stable to axisymmetric perturbations, if the charge increases (angular momentum of the fluid aligned to the magnetic field) or decreases (angular momentum of the fluid anti-aligned to the direction of the magnetic field) outwards.
    
    \paragraph{Case V: Constant angular velocity, charged central object.}
    In this case, the spherical coordinate system is used. In this coordinate system, the electric potential arising from the charge of the central object  is given by:
    \begin{equation}
       E_a=\left(\frac{Q}{r^2},0,0\right)\; .
    \end{equation}
    We consider no magnetic field, therefore setting $A_a=0$. The two conditions for stable structures for a constant angular velocity distribution and under axisymmetric perturbations reduce in this set up to:
    \begin{align}
        \Tr(M_{ab})&=4\Omega^2+ \partial_r {\hat q} E_r \, r\Omega \; ,\\
        \det(M_{ab})&=2r\Omega^2\cos\theta\, E_r\left[r\cos\theta\, \partial_r\hat q-\sin\theta \, \partial_\theta\hat q\right]\; .
    \end{align}
    According to this result, a charged fluid structure can be considered stable to axisymmetric perturbations, if the charge increases (same charge as the BH) or decreases (opposite charge of the BH) outwards. Structures along the equatorial plane seem stable, as long as the charge does not significantly change along the angle $\theta$, or, if it does, decreases (same charge as the BH) or increases (opposite charge of the BH) towards the equatorial plane.
    
    \paragraph{Case VI: Constant angular velocity, charged central object, homogeneous magnetic field.}
    This time a homogeneous magnetic background field is assumed as well as a charged central object. 
    
    Assuming the following charge distribution is motivated by the former work \cite{trova2020}. In this paper bound charged fluid structures were found with a charge distribution of
    \begin{align}
        \hat q =& f(S), & S=&-\Phi_E+\Omega A_\phi\; 
    \end{align}
    in a general relativistic setting.
    
    The two conditions for stable structures in case of a  constant angular velocity distribution and under axisymmetric perturbations reduce in this set up to:
    \begin{align}
        \label{hbfC1}
        \begin{split}
        \Tr(M_{ab})=&\frac{g^{ab}\partial_a g_{\phi\phi}\partial_b g_{\phi\phi}}{g_{\phi\phi}}\left(\Omega+\hat q A^\phi\right)^2\\
        &+f'(S)g^{ab}\left(E_a+\Omega \partial_a A_\phi\right)\left(E_b+\Omega \partial_b A_\phi\right),
        \end{split}\\
        \det(M_{ab})=&f'(S)\frac{\left(\Omega+\hat q A^\phi\right)^2}{g_{\phi\phi}}\left(\varepsilon^{ab}\partial_a g_{\phi\phi} (E_b+\Omega \partial_b A_\phi)\right)^2.
        \label{hbfC2}
    \end{align}
    Both conditions are fulfilled, if $f'(S)\geq0$.
    
    This setup was discussed in the general relativistic case by \citet{kovar14}. The equatorial tori fulfill the stability conditions in contrast to the chosen setup for the polar clouds. This could be a hint, that the found polar cloud solutions are actually unstable with respect to axisymmetric perturbations. A general relativistic discussion is needed for a sound answer.
    
    \paragraph{Case VII: Magnetic dipole field}

    Employing the spherical coordinates $\{r, \theta, \phi\}$, the background magnetic field can be derived from the vector potential 
    \begin{align}
        A_\phi&=\frac{\mathcal{M}}{r} \sin^2\theta, & &\text{and } & A^\phi&=\frac{\mathcal{M}}{r^3}\, .
    \end{align}
    In addition, if the electric field vanish, $\Phi_E=0$, the two conditions for stable equilibrium structures under the axisymmetric perturbations reduce to
    \begin{align}
        \mathrm{Tr}\left(M_{ab}\right) =& \frac{1}{r^2\sin^2\theta}\Bigg[
        \left(\frac{2\ell}{r} + \hat{q}\partial_r A_\phi\right)
        \left(\partial_r\ell + \hat{q}\partial_r A_\phi\right) 
        \nonumber \\
        & + 
        \frac{1}{r^2}\left(2\ell\cot\theta + 
        \hat{q}\partial_\theta A_\phi\right)
        \left(\partial_\theta\ell + \hat{q}\partial_\theta A_\phi\right) +
        \nonumber \\ 
        & +
        \ell\left(\partial_r A_\phi \partial_r\hat{q} + 
        \frac{1}{r^2}\partial_\theta A_\phi\partial_\theta\hat{q}\right)
        \Bigg] 
        \label{eq:dpole1}
    \end{align}
    and
    \begin{align}
        \mathrm{det}\left(M_{ab}\right) =& \frac{2\ell^2}{r^4\sin^4\theta}\Bigg[
        \frac{1}{r}\partial_\theta A_\phi - 
        \left(\partial_r A_\phi\right)\cot\theta \Bigg]
        \nonumber \\
        & \times 
        \Bigg[\left(\partial_r\ell + \hat{q}\partial_r A_\phi\right)
        \partial_\theta\hat{q} +
        \nonumber \\ 
        & \quad\quad
        \left(\partial_\theta\ell + \hat{q}\partial_\theta A_\phi\right)
        \partial_r\hat{q}\Bigg].
        \label{eq:dpole2}
    \end{align}
    In the case of a rigid-body rotation. $\Omega = \ell/(r\sin\vartheta)^2 = \mathrm{const}$, a fluid equilibrium exists when the specific charge can be expressed as function of the vector potential only, that is when $\hat{q} = f(A_\phi)$. In that case, the stability conditions (\ref{eq:dpole1}) and (\ref{eq:dpole2}) coincide with Eqs.~(\ref{hbfC1}) and (\ref{hbfC2}) given above. Consequently, such configuration are therefore stable as long as $f^\prime(A_\phi)>0$.

\section{Discussion}
    {
    More general equilibrium configurations, with specific angular momentum constant on cylinders $r\sin\vartheta = \mathrm{const}$ have been studied by Slan\'{y} et al \cite{Slan2013}. In particular, they consider angular momentum distribution given by a power-law functions of the form
    \begin{equation}
        \ell = K_2\left(r\sin\theta\right)^{K_1 + 1}
        \label{eq:slan-l}
    \end{equation}
    with $K_1$ and $K_2$ being constant parameters ($K_1=\pm1$ correspond to constant angular velocity and constant angular momentum, respectively; when $K_2 = \sqrt{GM}$ and $K_1 = -1/2$, Eq.~(\ref{eq:slan-l}) describes Keplerian flows). It is also assumed that the distribution of specific charge can be expressed as
    \begin{equation}
        \hat q =Cr^\lambda \sin(\theta)^\delta
        \label{eq:slan-q}
    \end{equation}
    with $C$, $\lambda$ and $\delta$ being yet another constants. The functions (\ref{eq:slan-l}) and (\ref{eq:slan-q}) describe equilibrium toroidal structures only when the parameters $K_1$, $\lambda$ and $\delta$ satisfy the condition of integrability. In order to reduce a large number of free parameters, Slan\'{y} et al \cite{Slan2013} consider four particular relations among the constants $K_1$, $\lambda$ and $\delta$ for which they are able to find bound fluid configurations. They referred to as the family I, II, III, and IV, respectively. In general, all four families allow equatorial tori, with pressure maxima located in the equatorial plane. These tori are similar to thick-disks solutions of purely hydrodynamic equations, modified by presence of Lorentz force. In addition, the families II -- IV permit also solutions where the Lorentz force pushes the pressure maxima out of the equatorial plane allowing off-equatorial ``levitating'' tori. As noted by Slan\'{y} et al \cite{Slan2013}, such configurations exist only for negative charge (i.e., when $\mathcal{M}q < 0$).

    In order to illustrate our results, we apply the stability criteria on both types of the solutions. We first investigate stability properties of equatorial tori corresponding to Family-I solutions. The relation between the parameters $\lambda$, $\delta$ and $K_1$ are (see Ref.~\cite{Slan2013})
    \begin{equation}
        \lambda = -\textstyle{\frac{3}{2}}\left(K_1 - 1\right),
        \quad
        \delta = 0.
    \end{equation}
    The equilibrium configurations are fully described by three independent parameters: the exponent in the angular momentum distribution $K_1$, the charge parameter of the torus $C$ and the maximal pressure $p_\mathrm{c}$ (i.e. \ the value of the pressure at the torus center). The Lorentz force, proportional to $C$, determines a departure of the fluid angular momentum at the center of the torus $\ell_\mathrm{c}$ from the local Keplerian value $\ell_\mathrm{K,c}$ and thus also the value of the parameter $K_2$. Here, we take the ratio $(\ell_\mathrm{c}/\ell_\mathrm{K,c})^2$ as a free parameter, instead of $C$, because of its immediate physical meaning. 
    
    Starting with slender tori ($p_\mathrm{c}\rightarrow 0$), in left-top panel of Fig.~\ref{fig2}, we plot possible configurations in the parameter space of $K_1$ vs $(\ell_\mathrm{c}/\ell_\mathrm{K,c})^2$. The horizontal line labeled as $q=0$ corresponds to uncharged hydrodynamic slender tori, for which $\ell_\mathrm{c} = \ell_\mathrm{K,c}$. The blue-shaded and red-shaded areas below and above this line correspond to allowed combinations of these two parameters for negatively- ($\mathcal{M}q < 0$) and positively- ($\mathcal{M}q > 0$) charged tori, respectively. These regions are bounded by dashed black lines, where the topology of the central point changes from maximum to saddle. Clearly only the former one gives reasonable solutions.
    
    Using expressions (\ref{eq:dpole1}) and (\ref{eq:dpole2}), we determine regions in the parameter space, where either $\mathrm{Tr}(M_{ab})$ or $\mathrm{det}(M_{ab})$ is negative. These areas correspond to hatched regions in the plot. According to the discussion in previous sections, we know that even slender tori corresponding to these parameters cannot be stable. Naturally, the same applies also to bigger tori with finite thickness, as their equi-pressure surfaces surround the ones of the slender tori. In other words, a stability of slender tori is a necessary condition for stability of tori with finite thickness, whose maximal pressure points are at the same locations. Therefore it follows that the only stable region in the parameter space left for possible stable configurations corresponds to negatively-charged tori with $K_1$ in the range $[-1, -0.5]$, where at least slender tori with negligible cross-section are stable. 
    
    We next examine tori of finite thickness with parameters corresponding to this region. The left bottom panel of Fig.~\ref{fig2} shows an example of the equi-pressure surfaces in the poloidal plane for the negatively charged torus with angular momentum at the center $\ell_\mathrm{c}^2 = 0.922\ell_\mathrm{K,c}^2$ and $K_1=-0.6$. The region of negative determinant $\mathrm{det}(M_{ab})$ is shown in the plot as a hatched area, the trace is positive everywhere. We observe that although the center of the torus lies in the stable area, our stability conditions put a serious constrain on the possible radial extend of the torus.        

    Next, we investigate stability of the off-equatorial (``levitating'') tori. We first concentrate on the Family-II solutions fully determined by three independent parameters: the radial angular momentum distribution $K_1$, height of the torus maximal pressure point above the equatorial plane $\cos\theta_\mathrm{c}$ and the central pressure $p_\mathrm{c}$. The family II corresponds to the charge distributions of the form 
    \begin{equation}
        \hat{q}=Cr^{3/2}(\sin\theta)^{-3K_1}
    \end{equation} 
    (see Ref.~\cite{Slan2013}) and the force balance at the torus center determines values of the two remaining constants $K_2$ and $C$. Again we start by discussing stability of slender tori in the limit $p_\mathrm{c}\rightarrow 0$. The center of the torus corresponds to the pressure maximum when $K_1<-1/2$, independently of the height of the torus above the equatorial plane. In addition, both $K_1$ and $\cos\theta_\mathrm{c}$ determine the stability of the torus center. The regions where the trace and determinant of the matrix $M_{ab}$ are negative are shown in the top right panel of Fig.~\ref{fig2} as differently hatched areas. Clearly, no combination of $K_1$ and $\cos\theta_\mathrm{c}$ gives a stable configuration. Consequently, neither tori of finite thickness can be stable. In the bottom right panel of Fig.~\ref{fig2} this finding is illustrated by an example of the structure of equi-pressure surfaces for $K_1=-2.25$ and $\theta_\mathrm{c} = 1.1$. The regions of negative trace and determinant of the matrix $M_{ab}$ are shown in the plot by differently hatched areas, showing that even the maximal pressure points of the toroidal structures are located well inside the unstable areas.

    These finding brinks up a natural question: Do \textit{stable} charged off-equatorial tori exist at all? To answer this question, we decide to go beyond the original analysis of Slan\'{y} et al. and let both $K_1$ and $\lambda$ to be independent parameters. We examine stability of slender off-equatorial tori at arbitrary height ($\sin^2\theta_\mathrm{c}\leq 1$). The result is shown in Fig.~\ref{fig3}. The blue-shaded area corresponds to parameters giving off-equatorial toroidal configurations. The families II, III and IV introduced by Slan\'{y} et al. \cite{Slan2013} correspond to the three color lines indicated in the plot. We also hatched the region where the trace or determinant of the stability matrix $M_{ab}$ is negative in full range of height, $0\leq\theta_\mathrm{c}\leq\pi$. Clearly, even with this more general approach, no combination of $K_1$, $\lambda$ and $\theta_\mathrm{c}$ may give a stable slender torus and our question thus has a negative answer. The same necessarily applies to tori of finite thickness, as for their stability, a stability of slender tori with the same $K_1$, $\lambda$ and $\theta_\mathrm{c}$ is necessary condition. 
    
    Finally, let us remark that the construction of off-equatorial tori can be still further generalized by allowing for different powers of $r$ and $\sin\theta_\mathrm{c}$ in the prescribed angular momentum distribution (\ref{eq:slan-l}). It is likely that for some particular values of these exponents a stable toroidal configuration will exist. Deeper explorations of this possibility is however behind the scope of this paper, where we mainly intent to illustrate power of our stability criteria. We wish to return to this question in our future work. 
    }
    \begin{figure}[tbh!]
        \centering
        \includegraphics[width=0.5\textwidth]{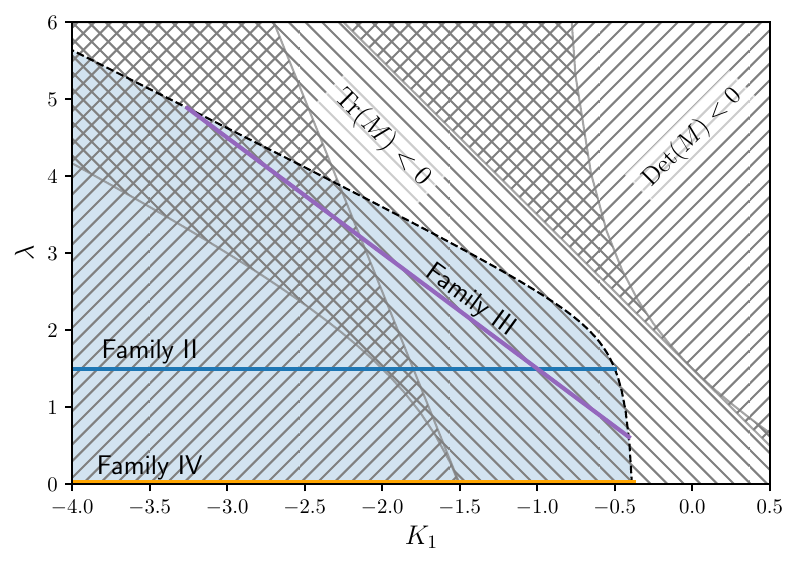}
        \caption{Three families (II, III and IV) also allow for {\em off-equatorial} tori in the regions of $K_1$--$\lambda$ parameter plane shown in the graph. Interestingly, none of these equilibrium configurations satisfy the stability criterion introduced in the present paper, as indicated by hatches covering the entire plane (see the main text for details). Each colored line corresponds to the range of $K_1$ values permitted by the equilibrium condition for three distinct families.}
        \label{fig3}
    \end{figure}

    \section{Conclusions}
    \label{conclusion}
    We investigated the linear stability of charged, non-conductive perfect fluid structures immersed in an electromagnetic background field concerning non-axisymmetric and axisymmetric perturbations in the Newtonian regime. In order to do so, we used the Lagrangian perturbation formalism introduced by \citet{Lebovitz1961}. 
    
    We found that the charge distribution within a non-conductive fluid cannot stabilize the general instability of ideal fluids with respect to non-axisymmetric perturbations, in agreement with the original argument by \citet{FriedmanSchutzb}. 
    
    In the case of axisymmetric perturbations, we found a generalization of the H{\o}iland conditions for stable fluid structures for our case of charged, non-conductive fluids. The generalized conditions were used to make specific statements concerning the stability of seven special cases of electromagnetic background fields and angular momentum distributions of a charged fluid structure. The seven cases were chosen either by their simplicity and direct generalization of a classical non-charged fluid structure (Polish doughnut with constant angular momentum) or because these setups had been discussed concerning the existence of bound charged fluid structures in previous works, and the question of stability of these structures needed to be answered.
    
    Charged fluid structures with a constant angular momentum distribution are found to be stable with respect to axisymmetric perturbations if the charge distribution within the structure increases outwards and gives rise to a repulsive Lorentz force everywhere.
    We could show that, while the found charged fluid structures discussed in \cite{kovar14} and some charged fluid structures of family I in \cite{Slan2013} are stable in the case of equatorial tori, the solutions for polar clouds appear to be unstable. 

    { Stability could also not be confirmed for the case of off-equatorial tori that had been discussed in \citet{Slan2013} with the magnetic dipole background field. On the other hand, modifying the assumed profile of the radial distribution of the charge density or the magnetic field structure will likely lead to stable configurations.}
    

\begin{acknowledgments}
We wish to acknowledge the continued support from the Czech Science Foundation EXPRO program (VK and JH, ref.\ 21-06825X) and the CTA-CZ research infrastructure of the Czech Ministry of Education, Youth and Sports (ref.\ LM2023047). AT and EH acknowledge the Deutsche Forschungsgemeinschaft (DFG, German Research Foundation) funded under the project number 420243324. Moreover, EH acknowledges the Deutsche Forschungsgemeinschaft (DFG, German Research Foundation) under Germany's Excellence Strategy - EXC-2123 QuantumFrontiers - 390837967. We further thank Shokoufe Faraji, Ji\v{r}\'i Kov\'a\v{r} and Petr Slan\'y for inspiring discussions.
\end{acknowledgments}

\appendix

\section{The (anti-)hermitian character of the operators in the perturbed charged Euler equation}
	\label{hermF}
	This section is dedicated to show that the operators $A_{ab}$, $B_{ab}$ and $C_{ab}$, occurring in Eq. \eqref{lp1}, are hermitian, anti-hermitian, and hermitian respectively. Since this has already been shown by \citet{FriedmanSchutza} in the uncharged fluid case, we will restrict the discussion here to those terms, which arise from the electromagnetic interaction of the charged fluid with the electromagnetic background filed.
	
	The operators are given by:
	\begin{align}
	    \eta^a A_{ab}\xi^b=& g_{ab}\eta^a\xi^b\; , \\
	    \eta^aB_{ab}\xi^b=& \eta^a(2 g_{ab}u^c\nabla_c -\hat q F_{ab})\xi^b\; ,\\
	    \begin{split}
	    \eta^aC_{ab}\xi^b=& \eta^a\left[\left(u^b\nabla_b\right)^2 \xi_a+\frac{1}{\rho}\left(\nabla_b\xi^b\nabla_a p-\nabla_a \xi^b \nabla_b p\right.\right.\\
	    &\left.-\nabla_a(\gamma p \nabla_b\xi^b)\right)+\xi^b\nabla_b\nabla_a\Phi_G+\hat q\xi^b\nabla_b\nabla_a\Phi_E\\
	    &\left.-\hat q \left(F_{ab}u^\gamma\nabla_\gamma\xi^b+\xi^b\nabla_bF_{a\gamma}u^\gamma\right)\right] \; .
	    \end{split}
	\end{align}
	Since $A_{ab}$ is hermitian, and the electromagnetic contribution in $B_{ab}$ obviously anti-hermitian, ($F_{ab}$ is anti-hermitian), only the electromagnetic terms in $C_{ab}$ need to be checked for their symmetry. These terms are
	\begin{equation}
	    \hat q\eta^a\xi^b\nabla_b\nabla_a\Phi_E-\hat q\eta^a \left(F_{ab}u^\gamma\nabla_\gamma\xi^b+\xi^b\nabla_bF_{a\gamma}u^\gamma\right)\, .
	\end{equation}
	Since covariant derivatives commute, the electric term ($\hat q\nabla_b\nabla_a\Phi_E$) has to be hermitian. The magnetic contribution $\hat q \left(F_{ab}u^\gamma\nabla_\gamma+\nabla_bF_{a\gamma}u^\gamma\right)$ can be rewritten as follows
	\begin{align}
	\begin{split}
	     q\eta^a &\left(F_{ab}u^\gamma\nabla_\gamma\xi^b+\xi^b\nabla_bF_{a\gamma}u^\gamma\right)\\
	     &\quad=\frac{ q}{2}\eta^a\xi^b\left[\nabla_a F_{bc}+\nabla_bF_{ac}\right]u^c\\
	     &\quad-\frac{q}{2}\left[ \eta^a F_{ab}u^c \nabla_c \xi^b+\xi^a F_{ab}u^c\nabla_c\eta^b\right]\\
	    &\quad-\frac{1}{2}\nabla_c\left(q u^c \eta^a F_{ab}\xi^b\right)\, .
	    \end{split}
	\end{align}
	{ The last term under the volume integral over the system vanishes because we demand the charge density at infinity to vanish, too. The two remaining terms are symmetric in exchanging $\eta$ and $\xi$. It is therefore shown, that the electromagnetic terms of $C_{ab}$ under an integral over the system volume are also hermitian.}

	\section{Second variation of the total energy}
	\label{svar}
	 The total energy of a charged, non-conductive fluid configuration is given by:
	\begin{align}
	\label{Etot}
	    E_{\mathrm{tot}}= T + U + W_G+W_E\; ,
	\end{align}
	where $T$ is the kinetic, $W_G=\int{\frac{1}{2}\Phi_G \rho\, dV}$ the mass potential, $W_E=\int{\frac{1}{2}\hat q \Phi_E \rho\, dV}$ the electric potential and $U$ the internal energy. The magnetic field does not contribute to the total energy of the system, since the Lorentz-force $F_{ab}u^b$
	is always perpendicular the the fluid motion. The given total energy therefore only has an additional electric potential term $W_E$ compared to the total energy of an uncharged, perfect fluid.
	
	Let us note that \citet{FriedmanSchutza} derived the variation of the total energy until second order for an uncharged, adiabatic, perfect fluid under the premise of mass conservation:
	\begin{widetext}
	\begin{align}
	\label{etotvnc}
	    \bm{\delta} E_{\mathrm{tot,nc}}=&\delta^1E_{\mathrm{tot,nc}} + \delta^2E_{\mathrm{tot,nc}} +\mathcal{O}(\xi^3)\\
	                       =&\int{\left[\rho \xi^i\left( u^j\nabla_j u_i + \frac{1}{\rho}\nabla_ip+\nabla_i \Phi_G \right)+  \rho u^\beta \Delta u_\beta \right] dV}\\
	                       &+\frac{1}{2}\int{ \rho\, \lvert \partial_t\xi\rvert^2 -  \rho\lvert u^\gamma \nabla_\gamma\xi\rvert^2+\gamma p \lvert\nabla_a\xi^a\rvert^2+2{\xi}^a \nabla_a p\nabla_b\xi^b+{\xi}^a\xi^b\rho\left[\nabla_b\nabla_a p+\nabla_b\nabla_a\Phi_G\right] dV}+\mathcal{O}(\xi^3)\; .
	\end{align}
	\end{widetext}
	In the charged case, the variation of the electric potential until second order needs to be added to the equation above. By taking into account, that the overall charge of the system is conserved (see Eq. \eqref{contch}, and $\Delta \hat q=0$) one gets analog to the variation of the gravitational potential:
	\begin{align}
	    \begin{split}
	    \mathrm{\delta} W_E=&\int{\hat q \Delta\left[\frac{1}{2}\Phi_E \rho\right]\, dV}\\
	    =&\int{\left[\hat q\rho \xi^a\nabla_a \Phi_E+\frac{1}{2}\hat q\rho\xi^a\xi^b\nabla_a\nabla_b\Phi_E\right]\,dV}\\
	    &+\mathcal{O}(\xi^3).
	    \end{split}
	\end{align}
	
	If the term $\rho u^\beta \Delta u_\beta$ is rewritten in the form $\rho u^\beta \Delta \hat u_\beta$ in the charged case the variation of the total energy up to second order will take the form:
	\begin{widetext}
	\begin{align}
	\label{vetotc}
	\begin{split}
	      \bm{\delta} E_{\mathrm{tot}}=&\int{\left[\rho \xi^i\left( u^j\nabla_j u_i + \frac{1}{\rho}\nabla_ip+\nabla_i \Phi_G +\hat q \nabla_i \Phi_E -\hat q F_{ij}u^j\right)+  \rho u^\beta \Delta \hat u_\beta \right] dV}\\
	      &+\frac{1}{2}\int \rho\, \lvert \partial_t\xi\rvert^2 -  \rho\lvert u^\gamma \nabla_\gamma\xi\rvert^2+\gamma p \lvert\nabla_a\xi^a\rvert^2+2{\xi}^a \nabla_a p\nabla_b\xi^b+{\xi}^a\xi^b\rho\left[\nabla_b\nabla_a p+\nabla_b\nabla_a\Phi_G+\hat q\nabla_b\nabla_a\Phi_E\right]\\
	      &-\rho \hat q \xi^a \left(F_{ab}u^\gamma\nabla_\gamma\xi^b+\xi^b\nabla_bF_{a\gamma}u^\gamma\right)dV+\mathcal{O}(\xi^3).
	\end{split}
	\end{align}
	\end{widetext}
	The first term in the first integral must vanish according to the Euler equation for charged fluids (see Eq. \eqref{feq}). Like in the uncharged case, the first order variation becomes zero, and the second order variation coincides with the canonical energy, if $u^\beta\Delta\hat u_\beta$ vanishes. In the case of a fluid in circular motion, this is the case, if the redefined angular momentum $\hat u_\phi=u_\phi+\hat q A_\phi$ is conserved. 
	
	Equation \eqref{vetotc} was derived by calculating the difference $\rho \hat q u^\beta \Delta A_\beta$ between $\rho u^\beta \Delta u_\beta$ and $\rho u^\beta \Delta \hat u_\beta$ to up to second order:
	\begin{align}
	    \Delta A_i = \Delta (g_{ib} A^b) =\Delta g_{ib} A^b + g_{ib} \Delta A^b + \Delta g_{ib} \Delta A^b\, , 
	\end{align}
	where 
	\begin{equation}
	    \Delta g_{ab} = \nabla_a\xi_b +\nabla_b\xi_a +\nabla_a\xi^\beta\nabla_b\xi_\beta +\mathcal{O}(\xi^3)
	\end{equation}
	and
	\begin{align}
	    \Delta A^i = \left(\delta_j^i -\nabla_j\xi^i +\nabla_j\xi^\beta\nabla_\beta\xi^i \right)\bar A^j -A^i+\mathcal{O}(\xi^3)
	\end{align}
	(see \cite{FriedmanSchutza}). Here, $\bar A^i$ is the perturbed vector field ${A'}^i$, framedragged along $\xi$. In our case, the magnetic vector potential is not affected by  the perturbations.
	
	Taylor expansion up to second order results in this case in:
	\begin{equation}
	    \bar A^i= A^i(x^a + \xi^a)= A^i + \xi^a \nabla_a A^i +\frac{1}{2}\xi^a\xi^b \nabla_a\nabla_b A^i +\mathcal{O}(\xi^3)\;. 
	\end{equation}
	Calculating all terms up to second order results in:
	\begin{equation}
	    \Delta A_i = \xi^a \nabla_a A_i + A_a \nabla_i \xi^a + \nabla_i \xi^a \xi^b \nabla_b A_a +\frac{1}{2} \xi^a\xi^b \nabla_a\nabla_b A_i\, .
	\end{equation}
	This leads to
	\begin{align}
	    \begin{split}
	   &\int{\rho \hat q u^\beta \Delta A_\beta\,dV}\\
	   &\quad=\int\hat q \rho u^\beta\left( \xi^a \nabla_a A_\beta + A_a \nabla_\beta \xi^a + \nabla_\beta \xi^a \xi^b \nabla_b A_a\right.\\
	   &\quad+\left.\frac{1}{2} \xi^a\xi^b \nabla_a\nabla_b A_\beta\right)\,dV
	   \end{split}\\
	   \begin{split}
	   &\quad= \int{\rho\,\xi^a\, \hat qF_{a\beta}u^\beta \, dV}\\
	   &\quad+\frac{1}{2}\int{\rho \hat q \xi^a \left( F_{ab} u^\gamma \nabla_\gamma \xi^b + \xi^b \nabla_bF_{a\gamma}u^\gamma\right) \, dV},
	   \end{split}
	\end{align}
	where we used $\nabla_i(\rho \hat q u^i=0)$ and assumed, that  the charge and mass density vanish at infinity. 
	By adding the variation of the electric potential to Eq. \eqref{etotvnc} and replacing $\rho u^\beta \Delta u_\beta$ by $\rho u^\beta \Delta \hat u_\beta-\rho \hat q u^\beta \Delta A_\beta$ finally leads to
	Eq. \eqref{vetotc}.


\bibliographystyle{unsrtnat}
\bibliography{Bibstability}

\end{document}